\documentclass[twocolumn,showpacs,amssymb,prb]{revtex4}
\usepackage{graphicx}% Include figure files
\usepackage{dcolumn}% Align table columns on decimal point
\usepackage{bm}% bold math

\begin{document}
\title{Unusual polymerization in the Li$_4$C$_{60}$ fulleride}
\author{M.~Ricc\`o}%
\email{Mauro.Ricco@fis.unipr.it}
\homepage{http://www.fis.unipr.it/~ricco/}
\author{T.~Shiroka}%
\author{M.~Belli}%
\author{D.~Pontiroli}%
\author{M.~Pagliari}%
\affiliation{Dipartimento di Fisica and Istituto Nazionale di Fisica della Materia,\\
Universit\`a di Parma, Parco Area delle Scienze 7/a, 43100 Parma, Italy}
\author{G.~Ruani}%
\author{D.~Palles}%
\altaffiliation[Permanent address: ]{Department of Physics, School of Applied Sciences, 
National Technical University of Athens, Heroon Polytechneiou 9 st., 15780 Zografos, Greece.}
\affiliation{Istituto ISMN-CNR, Via P.\ Gobetti 101, 40129 Bologna, Italy}
\author{S.~Margadonna}%
\affiliation{School of Chemistry, University of Edinburgh, West Mains Road, 
EH9 3JJ Edinburgh, UK}
\author{M.~Tomaselli}%
\affiliation{Laboratory of Physical Chemistry, ETH-H\"onggerberg, Pauli Strasse 10, 
CH-8093, Z\"urich, Switzerland}

\date{\today}%

\begin{abstract}
Li$_4$C$_{60}$, one of the best representatives of lithium intercalated 
fullerides, features a novel type of 2D polymerization.
Extensive investigations, including laboratory x-ray and synchrotron 
radiation diffraction, $^{13}$C NMR, MAS and Raman spectroscopy, 
show a monoclinic $I2/m$ structure, characterized by chains of 
[2+2]-cycloaddicted fullerenes, sideways connected by single C--C bonds. 
This leads to the formation of polymeric layers, whose insulating 
nature, deduced from the NMR and Raman spectra, denotes the complete 
localization of the electrons involved in the covalent bonds.
\end{abstract}

%61.10.Nz   X-ray diffraction
%81.05.Tp   Fullerenes and related materials -> Material Science
%61.48.+c   Fullerenes and fullerene-related materials -> Xrays - Structure
%82.35.Lr   Physical properties of polymers
%78 Optical properties, condensed-matter spectroscopy and other interactions 
%of radiation and particles with condensed matter 
%78.30.Na   Fullerenes and related materials -> Raman
%82.56.Ub   Structure determination with NMR -> Material Science 
%82.35.Lr   Physical properties of polymers  -> Material Science

%\keywords{Fullerene based superconductors, polaronic superconductors}%

\maketitle

\section{Introduction}
\label{sec:intro}
Extensive investigations of fullerene C$_{60}$, both in its pristine 
form as well as in its intercalated variants, have definitively 
confirmed the strong tendency toward polymerization of this molecule. 
Indeed, the original cubic van der Waals bonded fcc structure
of solid C$_{60}$ can easily be transformed into 1-, 2-, and even 
3-dimensionally connected polymers either by 
photo-excitation, \cite{regueiro95} or by high-temperature and 
high-pressure \cite{iwasa94,marques99} treatments.

The polymeric forms of C$_{60}$ have attracted considerable 
attention mainly because of the variety of crystal structures 
and their interesting magnetic, optical and mechanical properties.
Notably, ferromagnetism with Curie temperature above 500 K has 
recently been reported in the rhombohedral phase of the 2D-polymer. 
\cite{makarova01,wood02} On the other
hand, the intercalation of solid C$_{60}$ with electron donors like alkali
metals (A), has resulted in a wealth of fulleride salts with
stoichiometries A$_{x}$C$_{60}$, among which superconducting A$_{3}$C$_{60}$
have received extended attention. The intercalation of alkali metals itself
can, in few cases, induce the formation of 1D or 2D arrangements of
polymerized C$_{60}$. In particular, A$_{1}$C$_{60}$ (A = K, Rb, Cs) 
can display chains of fullerenes linked by ``double'' C--C bonds 
([2+2]-cycloaddition), \cite{winter98, chauvet94, pekker99} 
while 1D or 2D arrangements of singly bonded C$_{60}$ have been observed in 
Na$_{2}$RbC$_{60}$ \cite{bendele98,arcon99} or 
Na$_{4}$C$_{60}$. \cite{oszlany97,rezzouk02}

If compared with other better known alkali-doped fullerides, lithium 
intercalated compounds Li$_{x}$C$_{60}$ display quite different 
properties, which can be mainly ascribed to the small Li-ion dimensions 
and to the partial charge transfer from the Li-atoms.
The most relevant differences, addressed only in recent years, 
\cite{yasukawa01, maniwa01, tomaselli01a, cristofolini99, tomaselli01b}
concern the large doping range ($x=1\div30$), 
the lack of superconductivity and the formation 
of polymerized C$_{60}$ structures. \cite{tomaselli01a}

Lithium-doped fullerenes themselves behave differently depending 
on the amount of intercalated lithium. At high doping levels 
(typically for $x>7$) both diffraction \cite{cristofolini99} and multiple 
quantum NMR \cite{tomaselli01b} clearly show the formation of lithium 
clusters located in the largest interstices of the C$_{60}$ fcc lattice, 
with all the members of this subclass having similar properties.
However, as the lithium concentration decreases below $x = 7$, the volume 
occupied by the intercalants decreases, while the efficiency 
of charge transfer is expected to increase. Both these factors lead to
a reduction of the inter-fullerene distance, and if two facing carbon 
atoms get close enough, covalent bonds are formed giving rise to 
fullerene polymerization.

In this paper we extensively investigate Li$_{4}$C$_{60}$, which is 
the most representative member of this family by diffraction (laboratory 
and synchrotron) techniques, as well as $^{13}$C static NMR,  magic angle 
spinning (MAS) NMR and Raman spectroscopy. The polymerization seems to 
occur as a consequence of the ``positive'' chemical pressure exerted by 
Coulomb interaction, leading to %carbon-carbon bonding. 
a novel 2D polymerization pattern, characterized by the coexistence of single 
and ``double'' C--C bonds propagating along two orthogonal directions 
of the plane.
Detailed NMR lineshape analysis and Raman measurements, besides confirming
the structure, show beyond doubt the presence of an insulating phase, 
arising from the electron localization in the covalent bonds.

\section{Experimental Methods}
\label{sec:experiment}
Li$_x$C$_{60}$ samples $( 1 \leq \mathrm{x} \leq 6 )$ were prepared either 
by thermal decomposition of Li azide or by direct doping with lithium metal. 
In the former case a mixture of stoichiometric amounts of lithium azide
and C$_{60}$ powder was heated up to $\sim 510$ K under dynamical vacuum.
In the second case a pellet of metallic lithium and C$_{60}$ powder was 
heated up to $\sim 540$ K and, after several regrinding and pelleting 
cycles, a homogeneously doped sample is eventually obtained.
Preliminary x-ray characterization showed that both methods yielded 
good quality samples having the same structure. 

Laboratory x-ray diffraction was performed with a Bruker D8 diffractometer
equipped with a double G\"obel mirror monochromator and a GADDS position 
sensitive detector (Cu K$\alpha$ radiation), whereas synchrotron diffraction 
experiments were performed on the ID31 beamline at ESRF (Grenoble).
In both cases the samples were sealed in 0.5 mm quartz capillaries.

An AMX 400 Bruker workstation, equipped for solid state NMR experiments,
was used for the static $^{13}$C NMR measurements. The spectra were 
collected employing a standard Hahn echo pulse sequence 
(90$^{\circ}$--$\tau$--180$^{\circ}$) with $\tau = 100$ $\mu$s and 
a recycle delay of 200 s.
Magic angle spinning (MAS) $^{13}$C NMR measurements were performed 
with a home-built instrument. In this case the Hahn echo spectra were 
taken using a $\tau$ delay equal to twice the rotor period at a spinning 
rate of 5 kHz and 8 kHz respectively, with a recycle delay of 10 s.

Micro-Raman spectra at room temperature were re\-cord\-ed in a 
back-scattering geometry using a Renishaw 1000 and a T64000 Jobin-Yvon 
spectrometers, both equipped with charge coupled device (CCD) 
cameras and microscope lenses with $\times$100 and $\times$50 magnification. 
In both cases the resolution was $\sim 1.0$ cm$^{-1}$ and the accuracy better 
than 0.5 cm$^{-1}$.
The excitation was provided by the 488.0 nm and 514.5 nm lines of an Ar$^+$ laser 
or the 632 nm line of a He-Ne laser. The laser beam was focused on the sample in 
a spot of $\sim 2$ $\mu$m in diameter ($\times$100 lens), while the power 
level was kept below 50 $\mu$W %$50\times10^{-6}$ W 
to avoid photo-degradation effects. 
In the adopted scattering geometry the polarization of the incident and 
scattered light were parallel to each other. Accumulation times of several 
hours were needed to compensate for the very low levels of scattered light 
due to the low laser power.

%Figure 1
\begin{figure}
\includegraphics[width=0.45\textwidth]{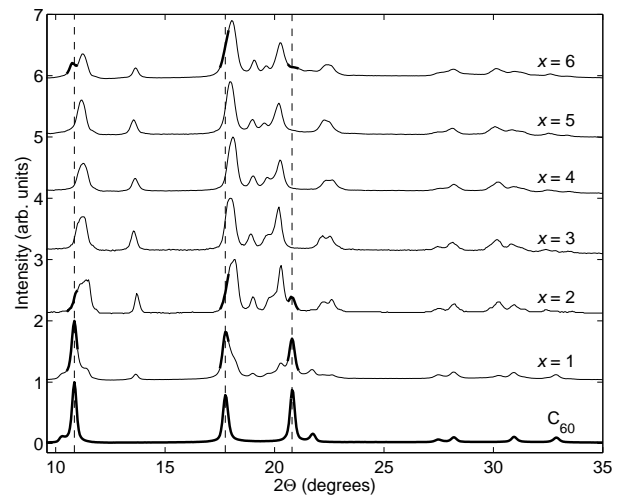}
\caption{\label{fig:LixC60}Room 
temperature Li$_x$C$_{60}$ powder diffraction patterns 
taken with Cu K$\alpha$ laboratory x-rays for $1 \leq x \leq 6$.
For $x=1$, 2, 6 there is a clear phase co-existence (bold lines), 
whereas for $3\leq x \leq 5$ the samples are homogeneous. 
For comparison the pristine C$_{60}$ diffraction profile and its 
main peak positions (dashed lines) are also shown.}
\end{figure}

%Figure 2
\begin{figure}
\includegraphics[width=0.45\textwidth]{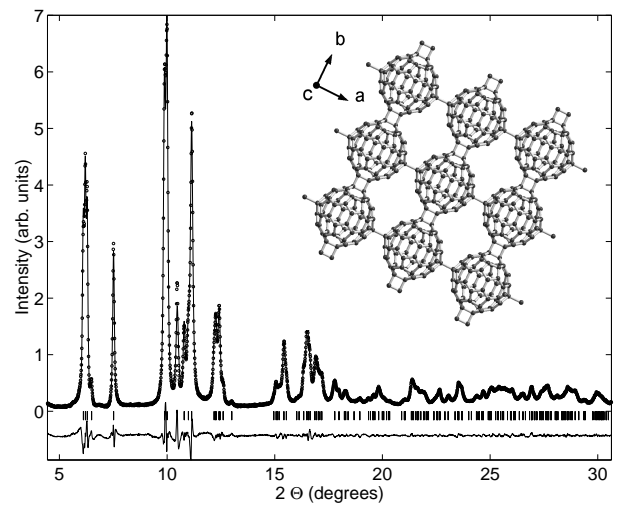}
\caption{\label{fig:li4c60refin}Measured (o) and calculated (solid line)
diffraction pattern of Li$_4$C$_{60}$ at 300 K. The 
lower solid line shows the difference profile and the ticks mark the 
reflection positions. Inset: the polymeric structure of Li$_4$C$_{60}$.}
\end{figure}

\section{Results and discussion}
\label{sec:results}

\subsection{Structural investigation}
\label{ssec:struct}
Laboratory x-ray diffraction measurements were performed on the whole 
Li$_x$C$_{60}$ series with $1 \leq x \leq 6$. The diffraction patterns show 
that for $ x \leq 2$ 
and $x = 6$ the samples consists of at least two phases, one of which is 
pristine C$_{60}$ which did not take part into the reaction (see Fig.~\ref{fig:LixC60}). 
On the other hand, for intermediate stoichiometries $3 \leq x \leq 5$, the compounds 
appear homogeneous and with the same structure. 
In fact, unlike other alkali doped fullerides (as e.g.\ K$_x$C$_{60}$), 
where the presence of line phases is observed, the structural 
properties of Li$_x$C$_{60}$ seem to depend weakly on stoichiometry.
Since Li$_{4}$C$_{60}$ is expected to be the most representative member of this 
family, detailed investigations were performed on samples having $x=4$.

Room temperature synchrotron radiation diffraction of 
Li$_{4}$C$_{60}$ \cite{margadonna04} (see Fig.~\ref{fig:li4c60refin}) 
shows that its structure is body centered monoclinic 
(space group \textit{I2/m}) with the following lattice parameters and angle: 
$a = 9.3267(3)$ \AA, $b = 9.0499(3)$ \AA, $c = 15.03289(1)$ \AA , 
$\beta = 90.949(3)^{\circ}$. The analysis with the Le Bail pattern 
decomposition technique provides $R_\mathrm{wp}= 4.24\%$ and 
$R_\mathrm{exp} = 1.53\%$. 
If center-to-center distances between the nearest C$_{60}$ units are 
considered, it turns out that along the \textit{b} direction the 
contact distance ($\sim 9.05$ \AA) is similar to that encountered in other 
polymerized A$_{1}$C$_{60}$, thus suggesting the presence of two bridging 
C--C bonds.

\begin{figure}
\includegraphics[width=0.45\textwidth]{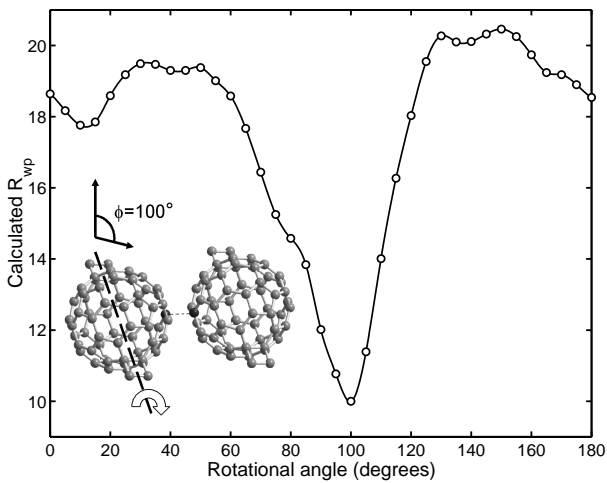}
\caption{\label{fig:rwp}Fit residuals $R_\mathrm{wp}$ as fullerene chains 
rotate along the \textit{b} axis. The minimum at $\sim 100^{\circ}$ corresponds 
to neighboring C atoms facing each other and giving rise to single bond 
polymerization (see inset).}
\end{figure}

The Rietveld refinement, starting from a structure derived from RbC$_{60}$, 
characterized by linear polymeric inter-fullerene chains directed along 
the \textit{b} axis,\cite{guerrero98,huq01} indicates that the fit quality 
strongly depends on the fullerene arrangement with respect to chain rotations 
along the [010] crystallographic directions. 
In particular, the fit residual $R_\mathrm{wp}$ shows a deep minimum 
when the angle $\phi$ with the initial direction becomes 
$\sim100^{\circ}$ (see inset in Fig.~\ref{fig:rwp}).

%Figure 3
At this angle, the arrangement of C$_{60}$ molecules is such that 
\textit{pairs of carbon atoms} on fullerenes belonging to neighboring 
chains are found very close ($\sim 2$ \AA) to each other. This strongly 
suggests that for a suitable orientation of the buckyballs not only the 
usual polymerization via the $[2+2]$-cycloaddition reaction along \textit{b} 
direction can take place, but also a connection through single C--C bonds 
along the \textit{a} axis is also possible. 
The minimum $R_\mathrm{wp}$ ($\mathrm{5.12 \%}$) in the final 
Rietveld analysis  was obtained by refining the positions 
of the whole C$_{60}$ atoms by the use of soft constrains.
It corresponds to a single interfullerene bond length of 1.75(2) \AA\ 
directed along the \textit{a} axis and to a bond inclination 
of $\sim 2^{\circ}$ with respect to the \textit{ab} plane. 

In an effort to localize the lithium atoms positions, the experimental 
diffraction pattern was investigated using Fourier analysis. 
The maxima of the electronic density map show that two lithium ions  
are located near the pseudo-tetrahedral site at (0.416, 0, 0.748),
whereas the other two Li$^{+}$ ions per fullerene are found in the
(0.023, 0, 0.380) and ($-0.023$, 0, 0.620) positions respectively, 
symmetrically displaced along the \textit{c} direction with respect 
to the center of the pseudo-octahedral site. In both cases the 
reported values refer to the average lithium positions, as confirmed 
by the high isotropic temperature factor ($B_{\mathrm{iso}} \sim 11$) 
and the smooth maxima in the Fourier map.

\subsection{NMR spectroscopy}
\label{ssec:NMR}
The uniqueness of the proposed Li$_{4}$C$_{60}$ structure requires 
a rigorous and systematic proof also by other independent investigations. 
Static and MAS (Magic Angle Spinning)$^{13}$C NMR measurements provided 
such a kind of independent evidence which fully confirmed the previous 
results.

\begin{figure}[floatfix,bht]
\includegraphics[width=0.45\textwidth]{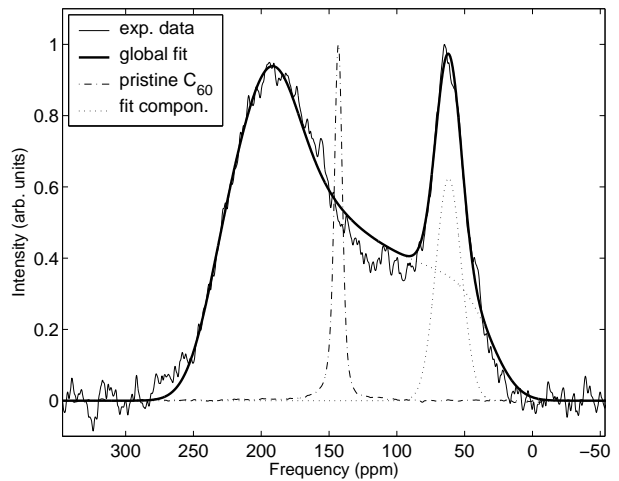}
\caption{\label{fig:static_nmr}$^{13}$C static NMR spectrum of Li$_4$C$_{60}$ 
at room temperature along with the global fit and its two components (a Gaussian 
function and a chemical shielding powder pattern). The narrow peak of pristine C$_{60}$ 
at 143 ppm is also shown.}
\end{figure}

The room temperature static spectrum of Li$_{4}$C$_{60}$, 
referenced to tetramethyl silane (TMS), is shown in Fig.~\ref{fig:static_nmr}
along with a reference spectrum of pristine C$_{60}$. 
The lineshape consists of a broad powder pattern, arising from 
the chemical shielding tensor, and a relatively narrow peak at lower 
frequencies which, at the intrinsic resolution of the static spectrum, 
does not exhibit any peculiar feature.
The broad pattern is located in a frequency domain commonly attributed
to \textit{sp}$^{2}$ hybridized carbon atoms, whereas the narrow peak falls
in the region of \textit{sp}$^{3}$ carbons.\cite{thier97}
%
%Figure 4

The lack of motional narrowing, typically observed in the room temperature
C$_{60}$ rotator phase, indicates that in the as-prepared Li$_{4}$C$_{60}$ 
the fullerene molecules are rotationally frozen.
At the same time, the presence of an \textit{sp}$^{3}$ peak at $\sim 60$ ppm 
confirms the formation of covalent bonds among C$_{60}$ units. A rough estimate 
of the number of carbon atoms involved in the bonds can be obtained by 
analyzing the integrated intensity of the different components of the spectrum. 
To this aim the spectra were fitted with a simple model, comprising a 
single chemical shift tensor pattern and an additional Gaussian peak. 
The ratio of the integrated intensities of the two fit components is 
respectively $1:8 (1)$, a result which is in good agreement with 
the expected $1:9$ ratio (6 carbon atoms involved in the 
bonds and 54 \textit{sp}$^{2}$-hybridized carbons per C$_{60}$ molecule).

A more detailed analysis of the \textit{sp}$^{2}$ component was not 
possible, since the contributions arising from the 14 inequivalent carbon 
atoms\cite{margadonna04} are not resolved in the static spectrum. 
Nevertheless, the tensor components can be considered as a rough average 
of the different contributions, enabling us to make useful comparisons 
with the carbon NMR spectra of other alkali-doped fullerides.
In particular, it is possible to obtain important clues about the 
nature of the global shift tensor $\bm{\delta}$, which represents the 
sum of the chemical shielding tensor $\bm{{\sigma}_{\mathrm{CS}}}$, 
arising from nucleus interaction with the orbital electrons, 
and the Knight shift tensor $\bm{K}$, originating from 
nucleus interaction with the delocalized conduction electrons. 
Thus, information about the electronic transport properties 
of Li$_{4}$C$_{60}$ can be directly extracted from the $^{13}$C 
NMR measurements.
%; this will lead us to conclude that the Knight shift
%tensor is negligible in Li$_4$C$_{60}$.

For an easier interpretation, the fitted $\bm{\delta}$-tensor 
components $[\delta_{xx}~ \delta_{yy}~ \delta_{zz}] = [237~ 191~ 31]$ 
can be conveniently separated into an isotropic part 
$\delta_{\mathrm{iso}} = (\delta_{xx}+\delta_{yy}+\delta_{zz})/3$ and an 
anisotropic traceless tensor 
$\bm{\delta_{\mathrm{aniso}}} = \bm{\delta} - \bm{\delta_{\mathrm{iso}}}$ 
(the convention adopted is 
$|\delta_{zz} - \delta_{\mathrm{iso}}| \geq |\delta_{yy} - \delta_{\mathrm{iso}}| \geq |\delta_{xx} - \delta_{\mathrm{iso}}|$). 
As far as the isotropic part is concerned, the value we find in the 
Li$_4$C$_{60}$ case, $\delta_{\mathrm{iso}}=$ +153 ppm, is quite close 
to the isotropic chemical shift value $\sigma_{\mathrm{iso}}=$ +143 ppm
of pristine C$_{60}$,\cite{tycko91} which, being the pure C$_{60}$ an 
insulator, coincides with its $\delta_{\mathrm{iso}}$.
%
%obviously does not contain any Knight shift contribution, being 
%the pure C$_{60}$ an insulator. 
In alkali-doped fullerenes, where the molecule is found as a 
C$_{60}^{n-}$ counter-ion, $\sigma_{\mathrm{iso}}$ is expected to 
increase slightly as a function of doping reaching $\sim 150$ ppm 
for $n = 3$,\cite{sato98} whereas $K_{\mathrm{iso}}$ for conducting 
fullerides is $\sim 40$ ppm.\cite{sato98,pennington96} The closeness of the
measured $\delta_{\mathrm{iso}}$ to that of pure C$_{60}$ implies the
lack of an \textit{isotropic} Knight shift in the static spectrum
of Li$_4$C$_{60}$, a conclusion that is fully consistent with the MAS 
results (\textit{vide infra}).
%
%thus making it a convincing proof of the insulating 
%nature of this polymer.

On the other hand, a contribution arising from the \textit{anisotropic} 
part of the Knight shift is still possible. %$\bm{K_{\mathrm{aniso}}}$. 
The traceless component of the fitted tensor is $[84~ 38~ -$$122]$,
implying an anisotropy parameter $\Omega=\delta_{zz}-\delta_{iso}=-$$122$ ppm.
The traceless tensor represents the sum of
the traceless chemical shielding tensor $\bm{\sigma_{\mathrm{aniso}}}$ and 
the traceless Knight shift tensor $\bm{K_{\mathrm{aniso}}}$, generally opposite 
in sign (we assume that the principal axes coincide for the two tensors). 
Typical values for chemical shielding anisotropy defined as 
$\Omega_{\mathrm{CS}}=\sigma_{zz}-\sigma_{iso}$
in alkali-metal doped fullerides fall around -110 ppm (and do not differ 
significantly from the chemical shielding 
anisotropy in pristine C$_{60}$),\cite{tycko91} while in metallic fullerides 
$\Omega_{\mathrm{K}}$ goes from +160 up to +200 ppm.\cite{sato98} 
From the comparison of our measurement with the values taken from the 
literature we conclude that also the Knight shift traceless tensor is 
negligible for Li$_4$C$_{60}$.
%
%Since the overall anisotropy $\Omega$ 
%in Li$_4$C$_{60}$ is much larger than $\sim 100$ ppm, expected in case 
%of metallic behaviour, we conclude that the anisotropy $\bm{\delta_{\mathrm{aniso}}}$ 
%is of purely chemical character.

%On the other hand, the traceless component of the fitted tensor is $[84~ 38~ -$$122]$, 
%implying an anisotropy parameter $\delta_{\mathrm{aniso}}=|\delta_{33}-\delta_{11}|=206$ ppm. 
%As a comparison we recall that the low-temperature chemical shielding 
%anisotropy in pristine C$_{60}$ is $\sim 180$ ppm.\cite{tycko91} In 
%alkali-doped fullerides the anisotropy parameter is not expected to 
%change significantly from this value. However, in metallic fullerides 
%the overall spectrum anisotropy is smaller, since the traceless 
%components of the chemical shift and the Knight shift tensors are of opposite sign. 
%Typical values of $K_{\mathrm{aniso}}$ fall in the range of $250 \div 320$ ppm\cite{sato98}. 

The analysis of both the isotropic and anisotropic components of the 
\textit{sp}$^2$ part of the spectrum represents therefore a convincing proof 
of the insulating nature of the Li$_{4}$C$_{60}$ polymer at room temperature.

Finally, we note that the complete absence of a sharp peak near $+143$ ppm, 
even with long recycle delays, implies the lack of unreacted C$_{60}$ in the 
sample, in agreement with the single phase evidenced by diffraction data.

Further insights come from the $^{13}$C MAS spectrum, where the anisotropy of 
the different interactions is averaged out, just like in NMR in solutions, and 
only the isotropic shifts are measured.

\begin{figure}[floatfix,t!]
\includegraphics[width=0.45\textwidth]{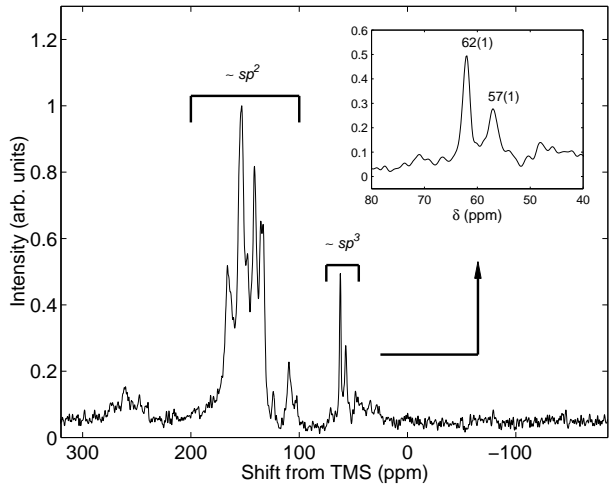}
\caption{\label{fig:mas_nmr}$^{13}$C MAS (8 kHz) spectrum at 300 K showing
the different intensities for $sp^2$ and $sp^3$ carbon atoms.
Insert: the two peaks in the $sp^3$ region arise from two
different types of C$_{60}$ polymerization.}
\end{figure}

A MAS measurement on Li$_4$C$_{60}$ using an 8 kHz spinning rate is reported 
in Fig.~\ref{fig:mas_nmr}.
The spectrum consists of a rather broad \textit{sp}$^2$ multiplet, spanning 
the region from $\sim 85$ MHz up to $\sim 210$ MHz, and two well resolved 
peaks in the \textit{sp}$^3$ hybridized carbon region, at +57 ppm and +62 ppm 
respectively.
By comparing the present spectrum with that obtained at a lower spinning 
rate (5 kHz, not shown here) allowed us to identify the satellite lines of 
the \textit{sp}$^2$ multiplet with the region which includes the small peaks 
from $\sim 210$ ppm to $\sim 300$ ppm and the broad basis below the 
two \textit{sp}$^3$ peaks.
On the other hand, the satellite lines originating from the \textit{sp}$^3$ 
peaks had a negligible amplitude and were not considered.

The relatively close spacing of the \textit{sp}$^2$ multiplet components, along with the low 
spinning rate imposed by geometrical constraints, prevented us from unambiguously assigning 
the \textit{sp}$^{2}$ peaks to each of the 14 structurally inequivalent \textit{sp}$^{2}$ 
carbon atoms; nevertheless some interesting conclusions can be drawn.
%A comparison between spectra recorded at 5 kHz and 8 kHz spinning rates (the
%latter is reported in Fig.~\ref{fig:mas_nmr}) points out that at 8 kHz the
%satellite lines have negligible amplitude. Thus, it is possible to clearly
%distinguish a variety of peaks in the \textit{sp}$^{2}$ carbons region,
%around the position of the motionally narrowed C$_{60}$ peak, and two
%distinct peaks at +62 ppm and +57 ppm. These two facts can be interpreted as follows: 

%Figure 5

The whole spectrum can be interpreted as follows:
the polymerization induces a distortion of the buckyballs, thus
different \textit{sp}$^{2}$ carbon atoms are characterized by slightly
different isotropic chemical shifts, clearly visible in the MAS spectrum. 
On the other hand, the two different peaks in the \textit{sp}$^{3}$ carbon 
region are reminiscent of the two different kinds of covalent bonds, which we 
can easily identify with the single bond along the \textit{a} direction and 
the four-membered carbon ring, analogous to a ``double'' bond, along the 
\textit{b} direction. Indeed, the integrated intensity ratio of these peaks 
is approximately $1:2$, in agreement with the number of carbon atoms involved 
in the bonds. Moreover, the lower frequency of the smaller \textit{sp}$^{3}$-like 
peak reflects its single bond nature, since in this case the more spherical 
hybridization acts as a better shield to the external field.

As a conclusion, both static and MAS $^{13}$C NMR data fully support not 
only the presence of polymerization in Li$_{4}$C$_{60}$, but also the complex 
and unusual bonding motif among the buckyballs.
In addition to carbon NMR we performed also $^{7}$Li static 
NMR measurements.\cite{ricco03}
The room temperature spectrum (not shown here) consisted of a single peak with 
an isotropic shift of approximately $-1$ ppm with respect to a saturated 
LiCl:aq reference solution. The nutation angle analysis led to the conclusion 
that it is a motionally narrowed peak, confirming once more the rather 
high mobility of Li atoms at room temperature and the fact that
x-ray diffraction can only identify their average positions.

\subsection{Raman spectroscopy}
\label{ssec:Raman}
The presence in fullerenes of ten Raman active modes (the $A_{\mathrm{g}}$ 
and $H_{\mathrm{g}}$ modes) makes Raman spectroscopy one of the preferred 
tools to study their dynamical and structural properties. 
\cite{kuzmany99,kuzmany04}
In a crystal, however, the selection rules derived from the high icosahedral 
symmetry of C$_{60}$ are somewhat relaxed, and several of the previously 
forbidden modes acquire Raman activity. It is on some of these `new' lines, 
which have become fingerprints of solid state phenomena like polymerization 
and dimerization, where we focus our attention.

In the Raman spectra of undoped polymeric C$_{60}$, the double peak in the 
region 940--980 cm$^{-1}$ has been assigned to the stretching vibrations of 
the four carbon atoms participating in the bridging [2+2] bonds between 
neighboring C$_{60}$ cages and is considered as a signature of polymeric 
C$_{60}$.\cite{adams95} Indeed, it is present both in the tetragonal\cite{wagberg02} 
(946 and 974 cm$^{-1}$, $\Delta\omega= 28$ cm$^{-1}$) as well as 
rhombohedral\cite{makarova01} (959 and 978 cm$^{-1}$, 
$\Delta\omega = 19$ cm$^{-1}$) phases of C$_{60}$. 
Nevertheless, the surprising presence of similar peaks (966 and 
980 cm$^{-1}$, $\Delta\omega = 14$ cm$^{-1}$) also in the polymerized 
\textit{single} C--C bond Na$_4$C$_{60}$,\cite{wagberg02} with a monoclinic 
two-dimensionally linked structure, has put serious doubts on the exclusive
assignment of these peaks to the ring stretch vibrations. Currently they 
tend to be associated with the more generic presence of 
\textit{sp}$^3$-type intermolecular bonds.

\begin{figure}[floatfix,t!]
\includegraphics[width=0.45\textwidth]{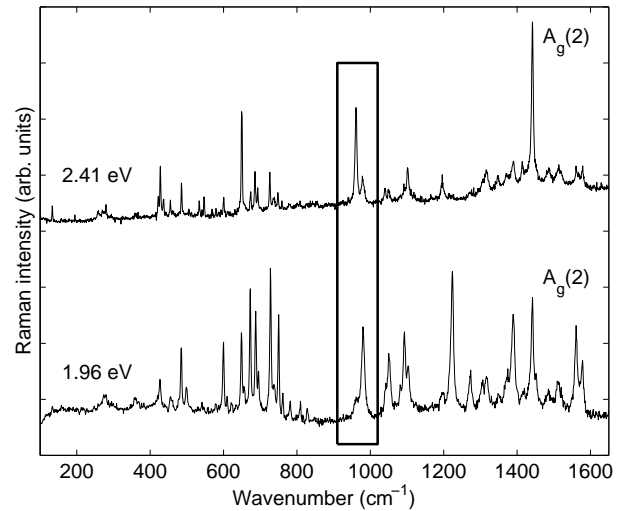}
\caption{\label{fig:raman}Room temperature micro-Raman spectra of 
Li$_4$C$_{60}$ in a parallel polarization scattering geometry and 
for two laser excitation energies: 2.41 eV and 1.96 eV. The two peaks
arising from carbon polymerization are shown in the boxed area.}
%The typical size of the crystallites was $\sim 10 \times 10$ $\mu$m$^2$.}
\end{figure}

The Raman spectra of Li$_4$C$_{60}$, synthesized by direct Li metal 
doping, are shown in Fig.~\ref{fig:raman}.
%
%Figure 6
%
%In the Raman spectra of undoped polymeric C$_{60}$, the double peak in the 
%region 940--980 cm$^{-1}$ has been assigned to the stretching vibrations of 
%the four carbon atoms participating in the bridging [2+2] bonds between neighbor 
%C$_{60}$ cages and is considered as a signature of polymeric C$_{60}$: \cite{adams95} 
%it is present both in the tetragonal\cite{wagberg02} (946 and 974 cm$^{-1}$, 
%$\Delta\omega= 28$ cm$^{-1}$) and rhombohedral\cite{makarova01} (959 and 978 
%cm$^{-1}$, $\Delta\omega = 19$ cm$^{-1}$) C$_{60}$. 
%
As expected, the presence of polymerization \textit{sp}$^3$ bonds gives 
rise to both the characteristic peaks (961 cm$^{-1}$ and 979 cm$^{-1}$,
$\Delta\omega = 18$ cm$^{-1}$), with positions and splitting very similar 
to those of rhombohedral C$_{60}$. Differently from what is found in the 
Na$_4$C$_{60}$ case, though, where both peaks have similar amplitudes 
and their relative heights do not change with laser excitation energy, 
in our case the analogous peak heights depend strongly on laser wavelength 
(see boxed area in Fig.~\ref{fig:raman}). This can be easily rationalized
if the complete equivalence of carbon bonds in the first case, 
and their mixed nature in the latter, is taken into account.

%The fact that the energies and the splitting of these peaks are similar in the
%as-prepared Li$_4$C$_{60}$ and Na$_4$C$_{60}$, but the lineshape of the bands 
%in Li$_4$C$_{60}$ looks similar to the tetragonal C$_{60}$ polymer agrees 
%with the x-ray description. 

A further confirmation about the lowered symmetry of the mixed polymeric 
structure, due to the presence of single bonds in one direction and 
``double'' C--C bonds in the orthogonal direction, as well as to a possible 
crystal field distortion due to the intercalated Li atoms, 
is found in the increased complexity of the Raman spectrum.
Indeed, in the low symmetry crystal structure of Li$_4$C$_{60}$ at least 
41 discernible bands can be recognized, to be compared with e.g.\ the 
higher symmetry tetragonal polymeric C$_{60}$, with only $\sim 23$ 
discernible bands (excluding very weak ones).\cite{wagberg02}
Moreover, also three low energy modes (which are silent in the highly 
symmetric crystalline C$_{60}$) and attributed to intramolecular 
stretching are observed as well at 133, 158 and 195 cm$^{-1}$. 

%thus observe some eigenmodes that were silent in the high symmetry crystalline 
%C$_{60}$, but also splitting of degenerate modes as a result of lifting of 
%the degeneracy with respect to the parent phase (see Fig.~\ref{fig:raman}). 

The Raman spectra of Li$_4$C$_{60}$ are generally characterized by narrow 
peaks, indicating both an ordered structure and a negligible electron phonon 
interaction, as in the case of non metallic systems.\cite{friedl92} 
In particular, the band assigned to the A$_{\mathrm{g}}$(2) parent mode, 
located at 1442 cm$^{-1}$ (see Fig.~\ref{fig:raman}), is very narrow 
($< 6$ cm$^{-1}$), confirming the presence of a single phase fulleride. 

Since many of the Raman peaks of Li$_4$C$_{60}$ appear either 
in the tetragonal two-dimensional polymeric C$_{60}$ (characterized by 
``double'' C--C bonds) or in the monoclinic two-dimensional polymeric network 
of Na$_4$C$_{60}$ (characterized by single C--C bonds), we can infer the 
mixed nature of our sample, in agreement with the results from x-ray 
diffraction. We point out, however, that although the present Raman spectrum 
is practically identical to that of Li$_4$C$_{60}$ reported in 
Ref.~\onlinecite{wagberg04}, the latter was characterized as a 
two-dimensional ``\textit{double}'' bond polymer.

The energy dependence of the A$_{\mathrm{g}}$(2) mode of C$_{60}$ (located 
at 1469 cm$^{-1}$ for the pristine material) has been correlated approximately 
linearly with the amount of charge transferred to the C$_{60}$ cage for the 
most studied alkali fullerides, A$_x$C$_{60}$, A = K, Cs, Rb, where the charge 
transfer has been verified independently using other experimental techniques.
In this case an A$_{\mathrm{g}}$(2) mode softening by 6--7 cm$^{-1}$ per 
electron transfer has been measured by Raman scattering.\cite{duclos91} 

For polymeric C$_{60}$, on the other hand, the issue of charge transfer is 
complicated by the presence of covalent bonds between C$_{60}$ units, 
which also causes a downshift of the A$_{\mathrm{g}}$(2) mode energy. 
According to the hypothesis reported in Ref.~\onlinecite{wagberg02}, the 
A$_{\mathrm{g}}$(2) mode energy downshift attributable to cage polymerization 
is $\sim 2.5$ cm$^{-1}$ per single 
%(in the case of Na$_4$C$^{­}_{60}$ the charge transfer appears 3e$^-$) 
and $\sim 5.5$ cm$^{-1}$ per ``double'' C--C bond.
Under this assumption, the charge transferred to C$_{60}$ in the 
Li$_4$C$_{60}$ case should be $[1469 - (2 \times 5.5 +% 
2 \times 2.5) - 1442]/6 \approx 2$e$^-$, presumably arising from 
only a partial charge transfer.
It remains therefore unclear whether the values of the  A$_{\mathrm{g}}$(2) 
mode energy in Na$_4$C$_{60}$ and Li$_4$C$_{60}$ can indeed be explained by 
this `rule' or if there exists in the polymeric systems some other mechanism  
that supersedes this simple additive rule and makes the identical 
A$_{\mathrm{g}}$(2) mode energies a direct consequence of a similar 
charge transfer in both cases. 

It is known that for Li$_x$CsC$_{60}$\cite{kosaka99} (and in part also 
for Na$_x$C$_{60}$\cite{andreoni96}) only an incomplete charge transfer 
from the alkali to the C$_{60}$ molecule takes place, in contrast to what 
is observed in other alkali metal doped fullerides. 
%In the case of Li$_4$C$_{60}$ and Na$_4$C$_{60}$ the charge transfer 
%to C$_{60}$ appears partial again. 

In the case of Li$_4$C$_{60}$ the large mobility of the Li atoms at room 
temperature, suggested by NMR measurements, is not compatible with the 
existence of hybrid electronic states between Li and C$_{60}$. 
Notably, Raman measurements performed in the high temperature monomeric 
phase of Li$_4$C$_{60}$, indicate a C$_{60}^{-4}$ charge state.\cite{ricco04}

\section{Conclusions}
\label{sec:conclusions}
In summary, the combination of powder diffraction analysis, NMR and 
Raman spectroscopies allowed us to precisely identify the structure as 
well as to probe the electronic properties of Li$_4$C$_{60}$, the best 
representative of low-doped lithium fullerides. 
A novel 2D polymeric arrangement with a mixed interfullerene bonding motif 
was found at room temperature. The formation of [2+2]-cycloaddicted 
fullerene chains sideways connected with single C--C bonds, suggested by 
the diffraction, is fully confirmed by NMR and Raman spectroscopy. 
The former shows two distinct \textit{sp}$^3$ lines in the $^{13}$C MAS 
spectrum whose intensity ratio agrees with the number of carbon atoms 
involved in the bonds, while the latter shows the presence of a doublet 
in the 940--980 cm$^{-1}$ region, a well known fingerprint of fullerene 
polymers.
Both $^{13}$C and $^{7}$Li NMR show the absence of Knight shift as well as 
any paramagnetic shift, clearly indicating that this polymer is a diamagnetic 
insulator. 
A measurement of the amount of charge transferred to C$_{60}$ from the Raman 
$A_{\mathrm{g}}(2)$ mode shift (a widely used method in alkali fullerides) 
cannot be reliably extended to fullerene polymers, where molecules are 
deformed by the presence of intermolecular bonds. Empirical rules 
\cite{wagberg02} suggest in Li$_4$C$_{60}$ only a partial transfer of two 
electrons, although diffraction Fourier analysis locates Li ions too far 
from the fullerene molecules to allow the hybridization of the 2\textit{s} 
electron (as it was observed also in Li$_x$CsC$_{60}$ \cite{kosaka99}).
This puts some doubts on the general application of these rules to 
different C$_{60}$ polymer structures.

\begin{acknowledgments}
We thank Dr.\ C.\ Vignali of the Centro Interfacolt\`a Misure 
of Parma University for his valuable help with the static NMR experiments
and acknowledge the financial support by the national FIRB project
``New Materials and Mechanisms of Superconductivity in Fullerenes''.

D.P.\ acknowledges the financial support by the European Commission 
through the T.M.R.\ network no.\ ERBFMRX-CT97-0155 ``FULPROP''
and the Greek State Scholarships Foundation (I.K.Y.).

\end{acknowledgments}

\end{document}